\documentclass[11pt,leqno,fleqn]{article}
\usepackage{amsthm,amsmath,amssymb}
\usepackage{graphicx}
\usepackage{subfigure}

\usepackage[authoryear]{natbib}

\bibpunct{(}{)}{;}{a}{,}{,}
\RequirePackage[OT1]{fontenc}

\newtheorem{theorem}{Theorem}

\newtheorem{proposition}{Proposition}

\theoremstyle{remark}

\theoremstyle{definition}
\newtheorem{definition}{Definition}

\newenvironment{ISItext}
{\normalsize\rm\setlength{\parindent}{1cm}\setlength{\parskip}{0pt}}
{\vskip 12pt}

\begin{document}
\newcommand{\ISItitle}[1]{\vskip 0pt\setlength{\parindent}{0cm}\Large\textbf{#1}\vskip 12pt}
\newcommand{\ISIsubtitleA}[1]{\normalsize\rm\setlength{\parindent}{0cm}\textbf{#1}\vskip 12pt}
\newcommand{\ISIsubtitleB}[1]{\normalsize\rm\setlength{\parindent}{0cm}\textbf{#1}\vskip 12pt}
\newcommand{\ISIsubtitleFig}[1]{\normalsize\rm\setlength{\parindent}{0cm}
\textbf{\textit{#1}}\vskip 12pt}
\newcommand{\ISIauthname}[1]{\normalsize\rm\setlength{\parindent}{0cm}#1 \\}
\newcommand{\ISIauthaddr}[1]{\normalsize\rm\setlength{\parindent}{0cm}\it #1 \vskip 12pt}

\ISItitle{Adaptive nonparametric detection in cryo-electron microscopy}

\ISIauthname{Langovoy, Mikhail}
\ISIauthaddr{Max Planck Institute for Intelligent Systems, Department Empirical Inference\\
and Max Planck Institute for Developmental Biology\\
Spemannstrasse 38\\
D-72076  T\"{u}bingen, Germany\\
E-mail: langovoy@tuebingen.mpg.de}

\ISIauthname{Habeck, Michael}
\ISIauthaddr{Max Planck Institute for Intelligent Systems, Department Empirical Inference\\
and Max Planck Institute for Developmental Biology\\
Spemannstrasse 35\\
D-72076  T\"{u}bingen, Germany\\
E-mail: michael.habeck@tuebingen.mpg.de}

\ISIauthname{Sch\"{o}lkopf, Bernhard}
\ISIauthaddr{Max Planck Institute for Intelligent Systems, Department Empirical Inference\\
Spemannstrasse 38\\
D-72076  T\"{u}bingen, Germany\\
E-mail: bernhard.schoelkopf@tuebingen.mpg.de}

\begin{ISItext}

\section{Introduction}\label{Section1}
Cryo-electron microscopy (cryo-EM) is an emerging experimental method to characterize the structure of large biomolecular assemblies.
Single particle cryo-EM records 2D images (so-called micrographs) of projections of the three-dimensional particle, which need to be processed to obtain the three-dimensional reconstruction.
A crucial step in the reconstruction process is particle picking which involves detection of particles in noisy 2D micrographs with low signal-to-noise ratios of typically 1:10 or even lower. Typically, each picture contains a large number of particles, and particles have unknown irregular and nonconvex shapes.

\section{Statistical model}\label{Section_Model}

Suppose we have a noisy two-dimensional pixelized image. In the present paper we are interested in detection of objects that have an \emph{unknown} colour. This colour has to be different from the colour of the background. Mathematically, we formalize this as follows.



We have an $N \times N$ array of observations, i.e. we observe $N^2$ real numbers ${\{Y_{ij}\}}_{i,j=1}^N$. Denote the true value on the pixel $(i, j)$, $1 \leq i, j \leq N$, by $Im_{ij}$, and the corresponding noise by $\varepsilon_{ij}$. Suppose that the noise on the whole screen is i.i.d. with an \emph{unknown} distribution function $F$. According to the above,

\begin{equation}\label{1}
Y_{ij} = Im_{ij} +  \varepsilon_{ij}\,,
\end{equation}

\noindent where $1 \leq i, j \leq N$, and $\{ \varepsilon_{ij}\}$, $1 \leq i, j \leq N$ are i.i.d., and $Im_{ij}$ denotes the true value on pixel $(i, j)$. Assume additionally that for the noise

\begin{equation}\label{2}
\varepsilon_{ij} \sim F, \quad \mathbb{E}\, \varepsilon_{ij} = 0,  \quad Var\, \varepsilon_{ij} = \sigma^{2} < + \infty\,.
\end{equation}

Our next assumption is inspired by images from cryo-electron microscopy. We assume that pixels in all particles have color intensity $b$, where $b > a$. In other words,

\begin{equation}\label{3}
\left\{
           \begin{array}{ll}
           Im_{ij} =  b > a, & \hbox{if $(i,j)$ belongs to a particle;} \\
           Im_{ij} = a, & \hbox{if $(i,j)$ does not belong to any particle.}
           \end{array}
         \right.
\end{equation}

\noindent Here we assume that $a$ and $\sigma^2$ are also both unknown. Estimating $a$ is one of the problems that are solved in this paper. The particle intensity $b$ is also not known, and one of our goals is to construct a good estimate of $b$ as well. The difficulty is that a priori we do not know which pixels contain pure noise and which pixels actually belong to some particle. Moreover, locations, shapes and exact sizes of particles are assumed to be unknown. The number of particles is also unknown, and in this paper we make no probabilistic assumptions about this number or about the distribution of particle locations. Thus we consider the case of a fully nonparametric noise of unknown level.

We make now the following crucial assumption. Let there exist a square $K_0$ of size at least $\varphi_0 (N) \times \varphi_0 (N)$ pixels such that $K_0$ doesn't intersect with any particle, and $\lim_{N \rightarrow \infty} \varphi_0 (N) \,=\, \infty $.

\noindent Here we only require that somewhere, in between the particles, there is one single square of size at least $\varphi_0 (N) \times \varphi_0 (N)$, such that this square is completely filled with noise. There could be several squares of this kind, or there could be bigger squares filled with noise. Locations of noisy squares are unknown. At the first step, our spatial scan estimator is going to find at least one of those noisy patches.



If a pixel $(i, j)$ is white in the original image, denote the corresponding probability distribution of $Y_{ij}$ by $P_0$. For a black pixel $(i, j)$ we denote the corresponding distribution by $P_1$. For any set of pixels $K$, we denote its cardinality by $|K|$. The empirical mean $\overline{Y}_K$ of all the values in $K$ is

\begin{equation}\label{5}
\overline{Y}_K \,=\, \frac{\,1\,}{\,|K|\,} \, \sum_{(i, j) \,\in\, K} Y_{ij} \,.
\end{equation}

\noindent The total sum of values in $K$ is denoted as

\begin{equation}\label{6}
S_K \,=\, \sum_{(i, j) \,\in\, K} Y_{ij} \,.
\end{equation}

Consider now some square $K$ of size $\varphi_0 (N) \times \varphi_0 (N)$ pixels. Suppose that $K$ contains $S_1^{(K)}(N)$ pixels from all of the particles in the image. The remaining $\varphi_0^2 (N) - S_1^{(K)}(N)$ pixels of $K$ contain pure noise. Obviously, $ S_1^{(K)}(N) \,\leq\, \varphi_0^2 (N) $ and, by the definition of $K_0$, $ S_1^{(K_0)}(N) \,=\, 0 \,$. In \cite{langovoy_scan_estimators}, the following auxiliary statement was proved.

\begin{proposition}\label{Proposition_1}

 1) If for some $K$ we have $\lim_{N \rightarrow \infty} \, S_1^{(K)}(N)\,/\,\varphi_0^2 (N) \,=\, 0$, then $\overline{Y}_{K}$ is a consistent estimate of $a$.


\noindent 2) In case if $\lim_{N \rightarrow \infty} \, S_1^{(K)}(N)\,/\,\varphi_0^2 (N) \,\neq\, 0$, it happens that $\overline{Y}_{K}$ is \emph{not} a consistent estimate of $a$.

\end{proposition}

\noindent This proposition clarifies why a naive approach of estimating $a$ by simple averaging of values over the whole screen leads to a generally inconsistent estimate. The naive estimator is consistent only when a combined size of all particles is negligible compared to the size of the screen. On the other hand, we see that sometimes the averaging procedure is consistent, namely, when the averaging is done over some $K$ such that this square has relatively little intersection with particles. This observation is crucial for our construction of scan estimators.

\section{Spatial scan estimators}


%
%
%
%
%
%

\begin{definition}\label{Definition_Spatial_Scan_Estimator}

Let $\mathcal{K}_0$ be the collection of all $\varphi_0 (N) \times \varphi_0 (N)$  subsquares of the screen. Set

\begin{equation}
\widehat{K} \,=\, \arg\min_{K \subseteq \mathcal{K}_0} \{ S_K \} \,.
\end{equation}

\noindent We define a \emph{spatial scan estimator} (for the lower intensity) as

\begin{equation}
\widehat{a} \,:=\, {\varphi_0 (N)}^{-2} \sum_{v \in \widehat{K}} Y_v \,.
\end{equation}

\end{definition}

\noindent The spatial scan estimator for noisy patches can be computed via the following algorithm.

\begin{itemize}
\item Step 1. Calculate means over all $\varphi_0 (N) \times \varphi_0 (N)$  subsquares.

\item Step 2. Select a neighborhood $\widehat{K}$ with the smallest mean.

\item Step 3. Define $\widehat{a} \,:=\, {\varphi_0 (N)}^{-2}\sum_{v \in \widehat{K}} Y_v$.

\end{itemize}

\noindent Since the initial noisy image can be naturally viewed as a square lattice graph, where $k^2$-nearest neighbors correspond to a $k \times k$ subsquare on the screen, we see that the spatial scan estimator (based on a popular method of sliding windows) is a special case of the $k$-NN scan estimator defined in \cite{langovoy_scan_estimators}.

Assume that each particle covers a square of size at least $\varphi_1 (N) \times \varphi_1 (N)$ pixels, and

\begin{equation}\label{7}
\lim_{N \to\infty} \frac{\varphi_1 (N)}{\log N} = \infty \,.
\end{equation}

The above idea can be inverted to get a spatial scan estimator for particle intensities.


\begin{definition}\label{Definition_Spatial_Scan_Estimator_Particles}

Let $\mathcal{K}_1$ be the collection of all $\varphi_1 (N) \times \varphi_1 (N)$  subsquares of the screen. Set $\widehat{K} \,=\, \arg\max_{K \subseteq \mathcal{K}_1} \{ S_K \}$. We define a \emph{spatial scan estimator} (for the higher intensity) as

\begin{equation}
\widehat{b} \,:=\, {\varphi_1 (N)}^{-2} \sum_{v \in \widehat{K}} Y_v \,.
\end{equation}

\end{definition}

\section{Consistency for bounded noise}\label{Section_Bounded_Noise}

In order to establish consistency of our spatial scan estimator, and to derive its rates of convergence, we might impose some additional assumptions on our model. Suppose that for all $i$ and $j$

\begin{equation}\label{18}
    |\, \varepsilon_{ij} \,| \,\leq\, M \quad \textit{almost surely}\,.
\end{equation}

\noindent This assumption is realistic in practice. For example, in the cryo-electron microscopy one often has $M \,=\, 1$.

The following bound is useful in quantifying behavior of spatial scan estimators (see \cite{langovoy_scan_estimators} and \cite{langovoy_habeck_schoelkopf_JSM} for proofs and more details).


\begin{proposition}\label{Proposition_7}
Let $K_0$ be any square with side $\varphi_0 (N)$ completely filled with noise, and let $\mathcal{K}$ be any collection of squares, each of size $\varphi_0 (N) \times \varphi_0 (N)$. Define $C_1 \,=\, 3\, (b - a)^2$, $ C_2 \,=\, 12\, \sigma^2 $, $ C_3 \,=\, 4 M \cdot (b-a) $. Then

\begin{equation}\label{28}
    P\, (\,\textit{some}\,\, K \,\in\, \mathcal{K} \,\, \textit{is chosen over} \,\, K_0 \,) \,\leq\, \sum_{K \,\in\, \mathcal{K}} \, \exp \, \biggr( -\, \frac{\, C_1 \, {\bigr[S_1^{(K)}(N)\bigr]}^2 \,}{\, C_2 \,|\,K \setminus K_0\,|\, +\, C_3 \, S_1^{(K)}(N) \,} \biggr) \,.
\end{equation}

\end{proposition}

%


\begin{theorem}\label{Theorem_1}
Suppose that the noise satisfies the above assumptions. If $\lim_{N \rightarrow \infty} \, \varphi_0 (N) {(\,\log N\, )}^{-1/2} \,=\, \infty$. Then the spatial scan estimator (of lower intensities) $\widehat{a} \,=\, \overline{a}_{\widehat{K}}$ is a consistent estimate of $a$. If $\lim_{N \rightarrow \infty} \, \varphi_1 (N) {( \,\log N\, )}^{-1/2} \,=\, \infty$, then the spatial scan estimator (for higher intensities) $\widehat{b}$ is consistent for $b$.
\end{theorem}

What can be said now about the rate of convergence of $\widehat{a}$ to the true $a$? In case if $K_0$ would be known, $\overline{a}_{K_0}$ would in general be an asymptotically efficient estimate of $a$, and also a $\sqrt{\, |\, K_0 \,| \,}$-consistent ($= \,\varphi_0 (N)$-consistent) estimate. One might expect that the rate might be at least $\frac{\, \sqrt{\,\log N\,} \,}{\, \varphi_0 (N) \,}$. The additional $\sqrt{\,\log N\,}$-factor is often called a "price for adaptation" in the literature, and estimates that slow down by a logarithmic factor are called "adaptive" (in our case, adaptation is to the extra difficulty of not knowing whether each particular observation belongs to the noise or not). However, it can be shown that the spatial scan estimator actually achieves the parametric rate of convergence. This is a stronger type of adaptivity.

\section{Simultaneous detection of multiple particles}


The key idea of our particle detection approach is to threshold the noisy picture at a properly chosen level, and then to use percolation theory to do statistics on black and white clusters of the resulting binary image. More details about our approach can be found in \cite{langovoy_report_2009-035}, \cite{langovoy_report_Robust_Detection}, \cite{Langovoy_Wittich_Square}, \cite{langovoy_wittich_report_R}, \cite{langovoy_wittich_report_Realistic_Pictures}, \cite{Langovoy_Wittich_Randomized_Algorithms} and \cite{Langovoy_Wittich_Robust}. See also related recent work \cite{Arias-Castro_etal_Cluster_Detection} and \cite{Arias-Castro_Grimmett}.


We transform the observed noisy image $\{Y_{i,j}\}_{i,j=1}^{N}$ in the following way: for all $1 \leq i, j \leq N$,\par\smallskip

1. $\quad$ If $Y_{ij} \geq \theta (N)$, set $\overline{Y}_{ij} := 1$ (i.e., in the transformed picture the corresponding pixel is coloured black).\smallskip

2. $\quad$ If $Y_{ij} < \theta (N)$, set $\overline{Y}_{ij} := 0$ (i.e., in the transformed picture the corresponding pixel is coloured white).\smallskip

\noindent The above transformation is called \emph{thresholding at the level} $\theta (N)$. The resulting array $\{\overline{Y}_{i,j}\}_{i,j=1}^{N}$ is called a \emph{thresholded picture}.

%
%

One can think of pixels from $\{\overline{Y}_{i,j}\}_{i,j=1}^{N}$ as of colored vertices of a sublattice $G_N$ of some planar lattice. We add coloured edges to $G_N$ in the following way. If any two black vertices are neighbours in the underlying lattice, we connect these two vertices with a black edge. If any two white vertices are neighbours, we connect them with a white edge. The choice of underlying lattice is important: different definitions can lead to testing procedures with different properties, see \cite{langovoy_report_2009-035}, \cite{langovoy_davies_wittich} and \cite{langovoy_report_Robust_Detection}. The method becomes especially robust when we work with an $N \times N$ subset of the \emph{triangular} lattice $\mathbb{T}^2$. Our goal is to choose a threshold $\theta (N)$ such that $P_0 (\,Y_{ij} \geq \theta (N)\,) < p_{c}^{site}$, but $ p_{c}^{site} < P_1 (\,Y_{ij} \geq \theta (N)\, )$, where $p_{c}^{site}$ is the critical probability for site percolation on $\mathbb{T}^2$ (see \cite{Grimmett}). If this is the case, we will observe a so-called \emph{supercritical} percolation of black clusters within particles, and a \emph{subcritical} percolation of black clusters on the background. There will be a high probability of forming large black clusters in place of particles, but there will be only little black clusters in place of noise. The difference between the two regions is the main component in our image analysis method.

\noindent \textbf{Algorithm 1 (Detection).}

\begin{itemize}
\item
Step 0. Estimate $a$ and $b$ via the spatial scan estimator. Set $\theta (N) := ( \widehat{a} + \widehat{b} )/2$.

\item Step 1. Perform $\theta(N)-$thresholding of the noisy picture $\{Y_{i,j}\}_{i,j=1}^{N}$.

\item Step 2. Until all black clusters are found, run depth-first search on the graph $G_N$ of the $\theta(N)-$thresholded picture $\{{\overline{Y}}_{i,j}\}_{i,j=1}^{N}$

\item Step 3. When a black cluster of size $\varphi_1 (N) $ was found, report that there is a particle corresponding to this cluster.

\item Step 4. If no black cluster was larger than $\varphi_1 (N) $, report that there were no particles.
\end{itemize}



\noindent Suppose that the image contains $\pi (N)$ particles. Assume

\begin{equation}
\lim_{N \to\infty} \frac{\,\pi (N)\,}{\,\log N\,} = 0\,.
\end{equation}

Let $A$ be a set of pixels corresponding to some particle. Define $thresh (A) \,:=\, \{\overline{Y}_{ij} \,|\, Y_{ij} \in A\}$. For each single particle $A$, let us say that the particle is \emph{detected} by the Algorithm, if $thresh (A)$ contains a significant black cluster. For a collection of several particles, the individual significant clusters can possibly merge into bigger clusters. We allow for this in the detection theorem, and leave the task of particle separation for the future pattern recognition work.

\begin{theorem}\label{Theorem1}
Let the noise be symmetric and satisfy the above assumptions. Then

\begin{enumerate}

  \item If there were $\pi (N)$ particles in the picture, Algorithm 1 detects all of them with probability that tends to 1 as $N \rightarrow \infty$.

  \item If there were no particles in the picture, the probability that the Algorithm 1 falsely detects a particle, doesn't exceed $\exp(-C_2(\sigma) \varphi_{im} (N))$ for all $N > N(\sigma)$.
\end{enumerate}

\noindent The constants $C_2 > 0$ and $N(\sigma)\in\mathbb{N}$ depend only on $\sigma$ and not on particle shapes or exact sizes.

\end{theorem}

%
%
%

\section{Application to particle picking in cryo-electron microscopy}
We applied our method to micrographs measured in cryo-electron microscopy.
The results are shown in Figure \ref{fig:cryoem}.
A micrograph of GroEL was downloaded from a public repository for particle picking (http://ami.scripps.edu/redmine/projects/ami/wiki/GroEL\_dataset\_I).
Before running the scan estimators the original $2400 \times 2400$ was down-sampled twice to improve the signal to noise and normalized to a maximum intensity of 1.
Our scan estimators yield $a=0.319$ (using a window size of 65) and $b=0.453$ (using a window with side 9), resulting in the threshold $\theta=0.386$.
After running the percolation analysis on the thresholded image (shown in Figure \ref{subfig:binary}) we delete black clusters occupying less than 30 pixels.
The result of the filtering procedure is shown in Figure \ref{subfig:filtered}. For each of the remaining black clusters, the algorithm claims that there is a particle containing this cluster.


\end{ISItext}

%
%

%
\begin{figure}
\centering
\subfigure[]{
	\includegraphics[scale=0.25]{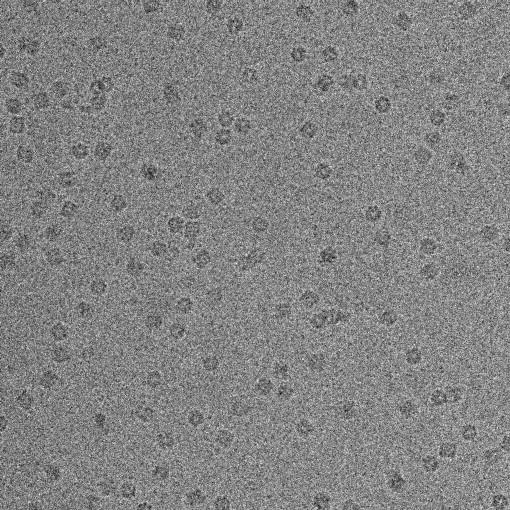}
	\label{subfig:original}
}
\subfigure[]{
	\includegraphics[scale=0.25]{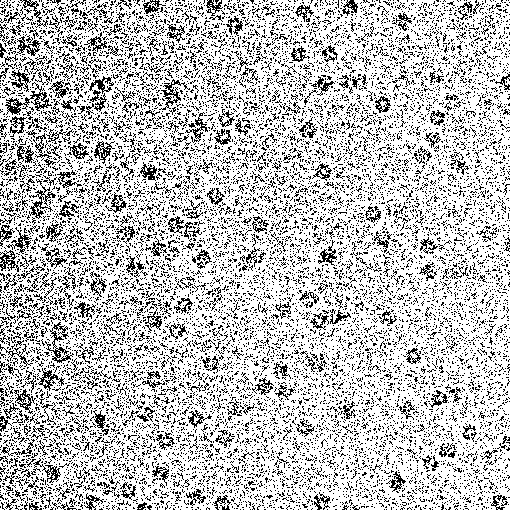}
	\label{subfig:binary}	
}
\subfigure[]{
	\includegraphics[scale=0.25]{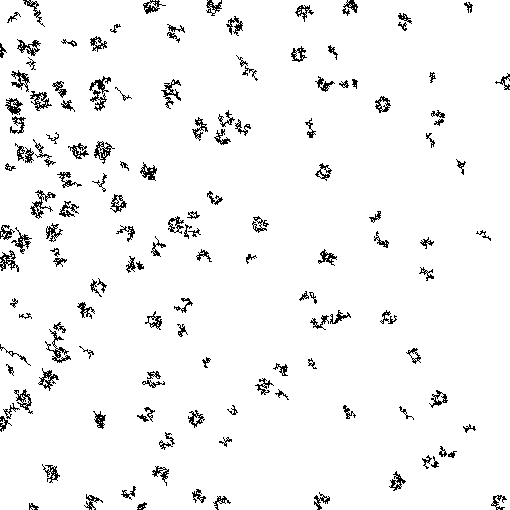}
	\label{subfig:filtered}	
}
\caption{%
Application to particle picking in cryo-EM.
(a): Original micrograph with particles of GroEL.
(b): Thresholded image based on spatial scan estimators of the noise and the signal.
(c): Thresholded image with random black clusters removed.
\label{fig:cryoem}}
\end{figure}
%

%

%
%

\bibliographystyle{plainnat}
\bibliography{papiere}

%

\end{document}